\documentstyle[11pt,newpasp,twoside,epsf]{article}
\def\edcomment#1{\iffalse\marginpar{\raggedright\sl#1\/}\else\relax\fi}
\reversemarginpar

\begin{document}
\title{The dark matter density problem in massive disk galaxies}
 \author{Benjamin J. Weiner}
\affil{UCO/Lick Observatory, University of California, Santa Cruz,
Santa Cruz, CA  95064,  USA}

\begin{abstract}
We discuss measurements of disk mass from non-circular streaming motions
of gas in the barred galaxies NGC 3095 and NGC 4123.  
In these galaxies with strong shocks and non-circular
motions, the inner regions must be disk-dominated to reproduce the
shocks.  This requires dark matter halos of low central density
and low concentration, compared to LCDM halo predictions.  In addition, 
the baryonic collapse to a disk should have compressed the halo and 
increased the dark matter density, which sharpens the 
disagreement. One possible resolution is a substantial amount of 
angular momentum transfer from disk to halo,
but this is not particularly attractive nor elegant.
\end{abstract}

\section{Introduction}

Schemes for galaxy formation which model the collapse of 
baryons within a dark matter halo can make useful predictions for
the structure of disk galaxies (e.g.\ Fall \& Efstathiou 1980,
Dalcanton et al.\ 1997, Mo et al.\ 1998).  Conversely,
measurements of disk galaxy properties should yield information
about the formation process.  For example, a success of the 
Fall \& Efstathiou picture is that it produces disks with the right
scalelengths, from the distribution of primordial
spins combined with the assumption that the baryons conserve angular
momentum during collapse.
Cosmological $N$-body and hydrodynamical modelers are now
capable of simulating disk galaxy formation with varying
degrees of succcess (e.g.\ Navarro \& Steinmetz 2000).
Simulating baryonic dissipation and star formation is still
quite uncertain, but $N$-body models make predictions for 
dark matter halo density and radial profile.  

Measurements of the radial distribution of luminous and dark mass in
disk galaxies are potentially an important test of disk formation
models.  However, there is a degeneracy between the luminous and
dark matter contributions to the  galaxy rotation curve, the 
``maximum disk'' problem (e.g.\ van Albada et al.\ 1985).
Several workers have approached this problem by studying dwarf and low
surface brightness galaxies, where the dark matter is expected
to dominate the rotation curve.  This line of inquiry has 
concentrated on the value of the inner slope of the DM halo density 
profile.  Whether the measurements show a $\rho =$ constant core
or a $\rho \propto r^{-1}$ cusp remains controversial, e.g.\ 
de Blok et al.\ (2001), Swaters et al. (2003), and several contributions
at this conference, and even $N$-body models have disagreed on
predictions for the inner slope.

I describe an alternative way to measure galaxy structure
and test formation models, using the internal non-circular 
kinematics of HSB barred galaxies to determine the 
contribution of luminous mass and constrain the properties of 
their dark halos.  These results do not constrain the ``core/cusp'' 
inner slope issue,
which in any case represents a disagreement over the distribution
of a very small percentage of dark halo mass, but constrain 
the dark matter mass within the optical disk of the galaxy.

\section{Barred galaxies, non-circular motions, and disk mass}

The velocity fields of barred galaxies contain additional
information beyond the overall rotation curve, which we may
use to break the disk-halo degeneracy.  Their gas velocity 
fields commonly have non-circular streaming motions and shocks
inside the bar.  These motions are caused by the elongated
potential of the stellar bar; the dark halo is usually presumed
to be dynamically hot and must be rounder than the bar.
The strength of the non-circular motions is dependent on
the ratio of disk to halo mass within the radius.

The gas flows contain shocks, so a full fluid-dynamical
treatmeant is necessary to model the gas velocity field.
Our goal is to constrain disk mass by comparing gas-dynamical
grid code simulations to high-resolution 2-D velocity fields from Fabry-Perot 
H$\alpha$ observations.  The full procedure is described in
Weiner et al. (2001).
We run many models for different values of the parameters
stellar disk $M/L_I$ and bar pattern speed $\Omega_p$; we
construct the potential using $I$-band photometry for
the stellar disk, and adding the dark halo needed to maintain
consistency with the overall rotation curve.  We then compare
the model velocity fields to observations inside the bar radius,
to constrain disk $M/L_I$ and bar $\Omega_p$.  Given the measurement
of disk $M/L$, the dark halo contribution to the rotation curve
is also measured.

\begin{figure}[tb]
\plottwo{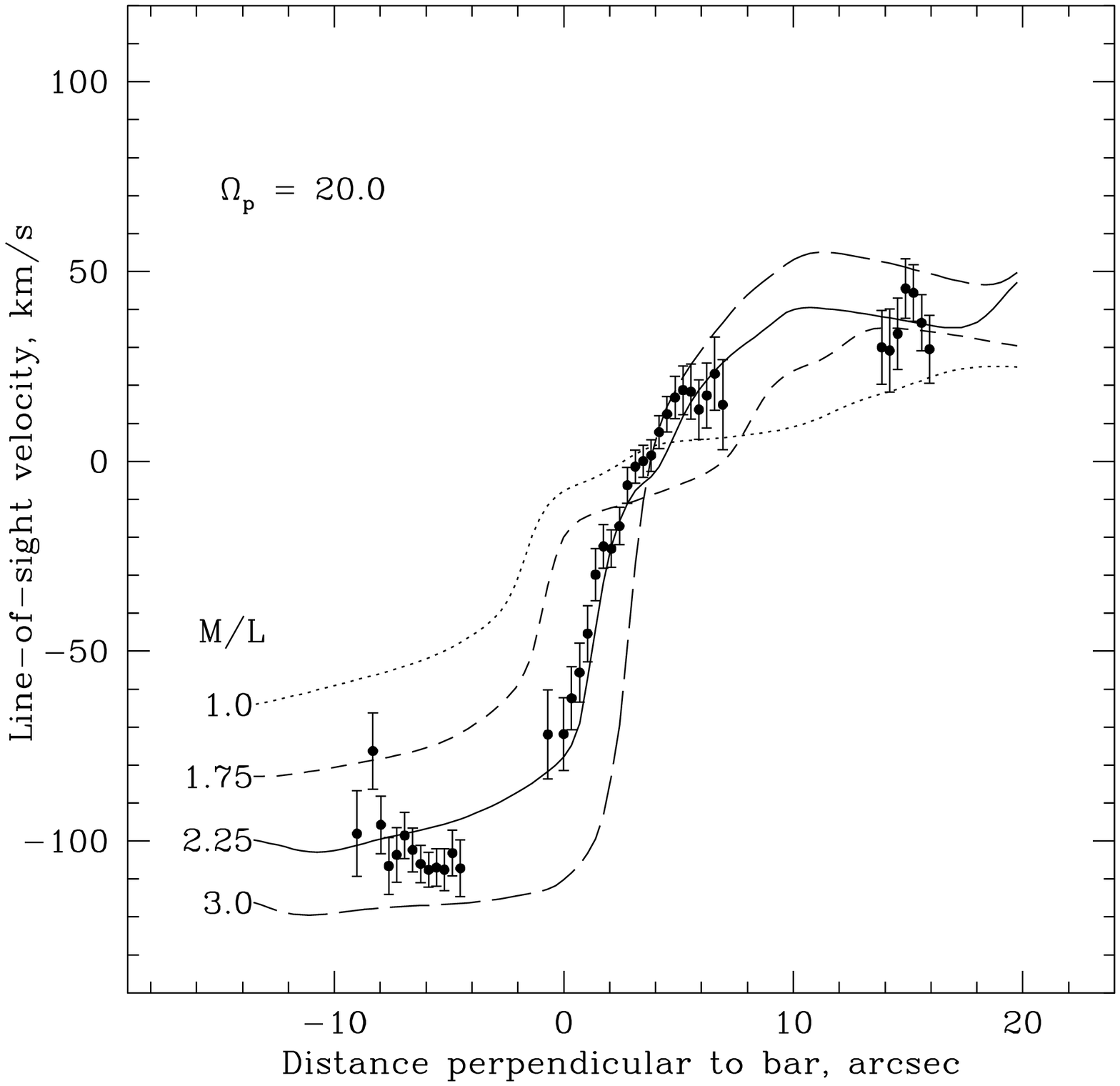}{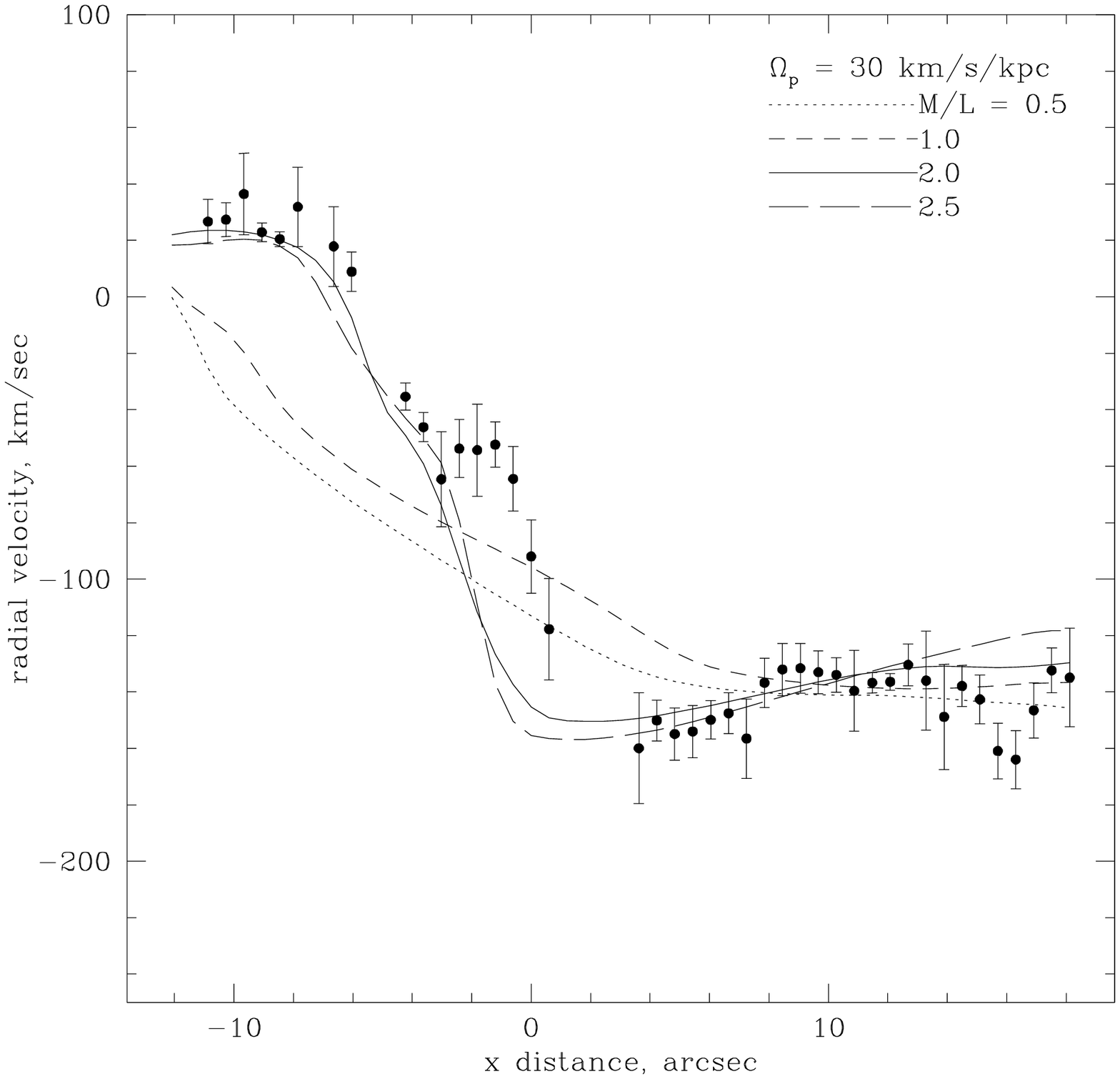}
\caption{Velocity jumps across the bar shocks in NGC 4123 (left)
and NGC 3095 (right), showing Fabry-Perot data (points) and 
models for various disk $M/L$ (lines).}
\end{figure}

Figure 1 shows, for the barred galaxies NGC 4123 and NGC 3095, 
a cut through the data and models 
which crosses the bar shock.  There is a sharp jump in velocity 
across the bar.  We plot several models to show the effect of 
varying disk $M/L$ while keeping pattern speed constant.  As
$M/L$ increases, the bar influence, non-circular motions, and 
shocks become stronger.  For $M/L_I \leq 1$ the shock is minimal.
A relatively high disk $M/L_I$ is needed to produce the observed 
shocks, 2.25 in NGC 4123 and 2.0 in NGC 3095.  The fit for
NGC 3095 is not perfect, but the sense is clear -- low mass
disks cannot make a strong enough shock.
Comparisons over the full 2-D velocity field inside
the bar bear this out (Weiner et al. 2001 for NGC 4123).
The comparison over the full range of model parameters 
also shows that a
fast-rotating bar with a high pattern speed $\Omega_p$ is required.

\section{Dark matter halo densities}

Given the measurement of disk $M/L$, we can return to the
rotation curve to determine the best fitting dark halo.
Figure 2 shows a fit to the HI rotation curve of NGC 4123
for the measured disk $M/L_I=2.25$.  The bar radius is 5 kpc, so the
disk $M/L$ is constrained only inside that radius by our
modeling.  However, at 5 kpc the disk contribution to the rotation curve
is already near its maximum; any falloff in $M/L$ outside
5 kpc has minimal effect.  This galaxy has $R_{25}=11$ kpc
and inside the optical disk the stars dominate the rotation
curve; it is very close to maximum disk.  
NGC 3095, which is larger in $V_{circ}$ and radius, is also 
close to maximum disk.

\begin{figure}[tb]
\plotfiddle{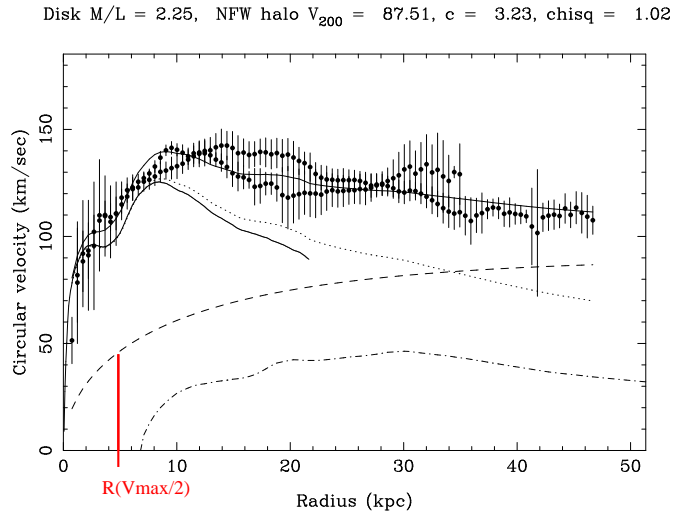}{6.5cm}{-90}{45}{45}{-180}{230}
\caption{NGC 4123
HI rotation curve with stellar disk (solid), HI gas (dot-dash),
stars+gas (dotted), and dark halo (dashed)
contributions for disk $M/L=2.25$.  The fiducial radius $R(V_{max})/2)$
of the dark halo is indicated.}
\end{figure}

Because these galaxies are dominated by stars in the center,
we cannot distinguish between different values of the inner
slope of the DM density profile.  However, we can put strong
constraints on the amount of dark matter within a given 
radius.  Alam et al. (2002) emphasize that it is 
easier to make a robust measure of the mean DM density 
within a fiducial radius than to measure the slope of a
density profile, since mean density 
$\Delta \propto M(<R)/R^3 \propto (V/R)^2$.  
They use the radius $R(V_{max}/2)$ where the
DM rotation curve reaches half-maximum velocity and compute
the mean enclosed DM density $\Delta(V_{max}/2)$
in units of the critical density.  

Figure 2 indicates the location
of $R(V_{max}/2)$ for NGC 4123, at $\sim 5$ kpc.  Because the 
stars dominate the inner rotation curve, $R(V_{max}/2)$ is
relatively large.  Equivalently the halo has a large core
or scale radius, and thus a low concentration $c_{NFW}=3.2$, lower than
predicted from LCDM simulations (Bullock et al. 2001, 
see Weiner et al. 2001).  The rotation curve fit can be tweaked
to make the DM halo curve rise more sharply, but the amount of DM
within 5-10 kpc is limited, and the DM must have a longer
scale length than the disk; there is not room for
much dark matter within a few scalelengths of the optical disk.

\begin{figure}[tb]
\plotfiddle{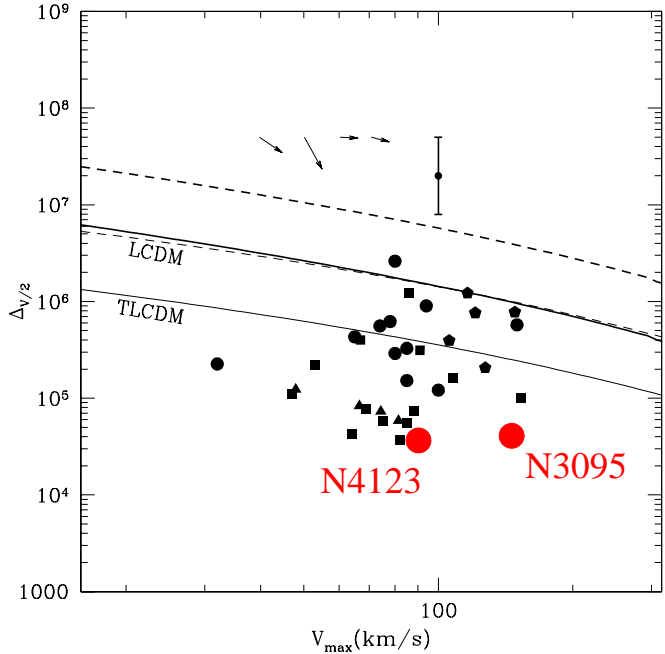}{8cm}{0}{65}{65}{-200}{-135}
\caption{$\Delta(V_{max}/2)$, mean DM density within $R(V_{max}/2)$,
veruss $V_{max}$, from Alam et al.\ (2002).  Small points are upper limits
from LSB galaxies, curves are LCDM and tilted LCDM models.  The
DM densities for NGC 4123 and NGC 3095 are overplotted (large points).  
These fall well below the LSB galaxies.}
\end{figure}

Figure 3 reproduces Figure 2 of Alam et al. (2002), showing
measurements of $\Delta(V_{max}/2)$ versus $V_{max}$ for a sample of
LSB galaxy 
rotation curves, and theoretical predictions from fitting DM halos in
LCDM and tilted LCDM simulations.  The LSB points are upper limits
since baryons may contribute slightly to the LSB rotation curves.
I have overplotted the mean densities $\Delta$ for the dark
halos of the barred galaxies NGC 4123 and NGC 3095.

The DM densities of the HSB barred galaxies are well below the 
LSB points and far below the LCDM and TLCDM predictions.
The disagreement is sharpened because the
predictions are for DM-only systems, pre-baryonic collapse,
while the HSB measurements are of post-collapse halos.
As baryons collapse into an angular momentum supported disk,
they move inward and should also draw the dark matter halo
inward.  This process is commonly modeled with the ``adiabatic 
compression'' assumption, which includes detailed conservation
of angular momentum for the dark matter (Blumenthal et al.\ 1986).  

It is extremely difficult to understand how the post-collapse halo
could be less concentrated than the pre-collapse state.  In fact,
when we attempted to reverse the adiabatic compression, running 
the collapse backwards for NGC 4123, we obtained at best 
an implausible, constant density initial state and at worst 
a simply unphysical initial density profile (Weiner et al. 2001).
We concluded that the assumptions of adiabatic compression must
be violated in order to produce the galaxy seen today.

\section{Conclusions}

In the two HSB barred galaxies we have modeled, the luminous
disk dominates the mass within several scalelengths, requiring a
low central density for the dark matter halo.  A similar conclusion
holds for the Milky Way based on its non-circular motions and on
microlensing (Gerhard, these proceedings).  The measured
{\it post-collapse} dark halo densities are lower than predicted
pre-collapse halo densities, even though the collapse process
should drive the DM density higher.

Two inexplicable galaxies are bad enough, but this is also 
a generic problem for HSB galaxies if the stellar disk 
dominates inside a few scale lengths.  The question of whether
disk galaxies are luminous or DM dominated has a long history.
Recently, based on the weak correlation of Tully-Fisher residuals with size
and surface brightness, Courteau \& Rix (1999) have argued that disk 
galaxies are very DM dominated, but our results, microlensing
studies of the Milky Way, and the ubiquity of $m=2$ spirals and bars
(Athanassoula et al.\ 1987) suggest that luminous matter must be
important inside a few scalelengths.  Since we don't yet understand
how the disk and halo are coupled in the formation process to make
the TF relation independent of surface brightness, it is not certain
what the TF residuals should show.
Kranz et al.\ (2003) have modeled spiral arm streaming
motions, and suggest that galaxies with $V_c>200$ km/sec are close
to maximum disk, while below that galaxies are DM dominated.  
(I speculate that DM-dominance depends on surface brightness; 
NGC 4123 is luminous matter dominated,
has a low $V_c=135$ km/sec, but is quite HSB.)

Are the disk-dominated NGC 4123 and NGC 3095 likely to be 
representative of HSB galaxies?  The $M/L_I$ ratios we find 
for NGC 4123 and NGC 3095 are reasonable for their colors, 
compared to the models of Bell \& de Jong (2001).
The galaxies are on the $I$-band Tully-Fisher relation 
(Giovanelli et al. 1997).  Barred and unbarred galaxies are
on the same TF relation and should have similar halo properties
(Courteau et al.\ 2003).  It seems unlikely that galaxies of
similar colors and surface brightnesses could have very
different $M/L$ and inhabit the same TF relation.

How can the low-density dark halos we find be reconciled with
the expectations of CDM?  It is possible to modify the initial
power spectrum to lower the power on small scales, but probably
not more than the TLCDM model shown in Figure 3.  Even that
does not explain why the post-collapse HSB galaxy halos are less 
dense than the pre-collapse predictions or LSB upper limits.  

Under the assumption of
adiabatic compression, the baryons always tend to drag the
DM inward, and it is difficult to keep the halo's scale length
much greater than the disk's.  Two potential mechanisms are
(1) blowout of baryons, which could loosen the binding of the DM,
and (2) transfer of angular momentum from baryons to DM.
It is not clear that blowout is effective in galaxies with
high $V_c$.  Angular momentum transfer is potentially very effective,
but unfortunately may upset the elegant paradigm for disk
formation developed since Fall \& Efstathiou (1980); for example,
if the baryons lose a significant amount of angular momentum,
disk sizes may no longer be predicted naturally.

In conclusion I suggest that we should not regard observations of
galaxy structure as merely tests of which, if any, CDM model is correct,
but as probes of the astrophysics of
galaxy formation.  Progress in this field may come as we move 
beyond comparing CDM-only halos to observed galaxies; real galaxies have
baryons, and we need a better understanding of what happens
during the collapse phase of disk formation.

\acknowledgments
I thank the American Astronomical Society for a travel grant
which allowed me to attend this conference.

\end{document}